\documentclass[11pt]{article}\topmargin=-0.7cm \textheight 222mm \textwidth=16.8cm\oddsidemargin=2mm 
\newcommand{\comm}[1]{}
\usepackage{amssymb}
\usepackage{amsmath}\usepackage{graphicx} \usepackage[numbers]{natbib}\usepackage{epsfig}
\def\citet{\cite}
\def\iindex{}
\def\xxxonly{\comm}
\def\xxxonly{ }
\def\noxxx{\comm}
\usepackage{times}\hfuzz=10pt 

 \usepackage{color}

\newtheorem{theorem}{Theorem}[section]

\newtheorem{proposition}[theorem]{Proposition}
\newtheorem{lemma}[theorem]{Lemma}
\newtheorem{corollary}[theorem]{Corollary}

\newtheorem{definition}[theorem]{Definition}
\newtheorem{remark}[theorem]{Remark}

\def\e{\varepsilon}

\def\defi{\stackrel{{\scriptscriptstyle \Delta}}{=}}

\def\OO{{\scriptscriptstyle O}}

\def\a{\alpha}
\def\d{\delta}
\def\o{\omega}
\def\O{\Omega}

\def\Y{{\cal Y}}
\def\F{{\cal F}}
\def\w{\widehat}
\def\Ind{{\mathbb{I}}}

\def\Re{{\rm Re\,\!}}

\def\R{{\bf R}}

\def\Z{{\cal Z}}
\def\ZZ{{\bf Z}}
\def\PP{{\cal P}}

\def\b{\beta}
\def\s{\delta}
\def\g{\gamma}
\def\C{{\bf C}}

\def\ww{\widetilde}
\def\X{{\cal X}}
\def\t{\theta}
\def\oo{\bar}
\def\s{\sigma}

\def\M{{\cal M}}

\def\T{{\mathbb{T}}}

\newcommand{\be}{\begin{equation}}
\newcommand{\ee}{\end{equation}}
\newcommand{\bd}{\begin{displaymath}}
\newcommand{\ed}{\end{displaymath}}
\newcommand{\ba}{\begin{array}{ll}}
\newcommand{\ea}{\end{array}}
\newcommand{\baa}{\begin{eqnarray}}
\newcommand{\eaa}{\end{eqnarray}}
\newcommand{\baaa}{\begin{eqnarray*}}
\newcommand{\eaaa}{\end{eqnarray*}}

\def\PP{{\cal P}}

\def\oo{\bar}

\def\a{\alpha}

\def\K{{\cal K}}
\def\yo{y}
\def\ew{\left(e^{i\o}\right)}
\def\BL{{\scriptscriptstyle BL}}
\def\BLO{L_{2}^{\BL,\O}(\R)}
\def\BLOO{L_{2}^{\BL,\pi/\tau}(\R)}
\def\T{{\mathbb{T}}}
\def\ZZ{{\mathbb{Z}}}

\def\TT{{\cal T}}

\def\HHH{{\rm H}}

\def\ee{\epsilon}

\def\dm{{\mnu}}
\def\SSS{{ subseq}\!}

\def\yy{y}
\def\qq{q}
\def\WW{\varrho}
\def\mnu{}
\title{On recovery of sequences from subsequences\xxxonly{\!\!\! : the case of non-periodic spectrum gaps}}
\author{
Nikolai Dokuchaev}

\begin{document}

 \vspace{-0.5cm}
\def\brea{}
\def\breakk{}
\def\break{}
\maketitle 
 \let\thefootnote\relax\footnote{The author is with  Department of
Mathematics and Statistics, Curtin University, GPO Box U1987, Perth,
Western Australia, 6845 (email N.Dokuchaev@curtin.edu.au). }
\begin{abstract}
\xxxonly{The paper investigates recoverability of  sequences from their periodic subsequences
and offers some modification of the approach suggested in papers  arXiv:1605.00414  	
and arXiv:1803.02233.
It is shown that there exists a class of sequences that is everywhere dense
in the class of all square-summable sequences and such that its members can be recovered from their periodic
subsequences. This recoverability is associated with certain spectrum degeneracy of a new kind.
 \noxxx{Application of this result allow to describe special classes of  continuous time band-limited functions
  allowing arbitrarily close approximation of  close
   to equidistant samples o
   featuring this degeneracy allows to establish recoverability of these functions from samples below the critical
     Nyquist rate.}}
     \noxxx{The paper investigates recoverability of sequences from their sparse decimated subsequences.
It is shown that  this recoverability is associated with certain spectrum degeneracy of a new kind defined via properties of subsequences, and that sequences of a general kind can be approximated by sequences with this feature.
  Application of this result  to equidistant samples of continuous time band-limited functions
   featuring this degeneracy allows to establish recoverability of these functions from sparse subsamples below the critical
     Nyquist rate.}

\end{abstract}
\section{Introduction}
\xxxonly{
The paper investigates  recoverability  of infinite sequences from their periodic
decimated subsequences
and offers some modification of the approach from  \cite{D16x} and \cite{D18x}.

 \noxxx{As is known, the sampling rate required to recover a continuous time function is defined by the classical Sampling Theorem
 also known as Whittaker-Shannon-Kotelnikov  theorem,
which is one of the most basic results in the theory of signal processing and information
science. Numerous  extensions of the sampling theorem  were obtained; see  the review of the literature
in \cite{D16x}.}

 It was shown in \cite{D16x} that a recoverability of a discrete time process from its subsequences is associated with certain "branching" spectrum degeneracy  based on periodic spectrum gaps on the unit circle $\T=\{z\in\C:\ |z|=1\}$ for Z-transforms, and that a sequences of a general kind can be approximated by sequences featuring this degeneracy. This degeneracy was described via representing the underlying process as a member of an ordered set  of processes  featuring periodic spectrum gaps and some common paths (a "branching" process).
 In \cite{D18x}, this result was  extended on spectrum degeneracy  of Z-transforms at  periodic systems of periodic isolated points on $\T$.
The present paper reformulated  the main  result from  \cite{D16x}  and related results from \cite{D18x} in a different way without using the branching structure and without using neither periodic spectrum gaps on $\T$  nor periodic
points of degeneracy on $\T$. This helped  to streamline the exposition.
}

\noxxx{

The paper investigates  recoverability of sequences from their decimated subsequences.
It appears that this recoverability is associated with certain spectrum degeneracy of a new kind, and that a
sequences of a general kind can be approximated by sequences featuring this degeneracy.
This opens some opportunities for sparse sampling of continuous time band-limited functions.
 As is known, the sampling rate required to recovery of  function is defined by the classical Sampling Theorem
 also known as
Nyquist-Shannon theorem, Nyquist-Shannon-Kotelnikov theorem,   Whittaker-Shannon-Kotelnikov theorem, Whittaker-Nyquist-Kotelnikov-Shannon theorem,
which is one of the most basic results in the theory of signal processing and information
science;  the result  was obtained independently by four  authors  \citet{Whit,Nyq,Kotel,Shannon}. The theorem states that
 a band-limited function can be uniquely recovered without error  from  a infinite two-sided equidistant
sampling sequence   taken with sufficient frequency:  the sampling rate must be at least twice the maximum frequency
present in the signal (the critical Nyquist rate). Respectively,  more frequent sampling is required for a larger  spectrum domain. If the sampling rate is preselected, then it is impossible to approximate a function of a general type by a band-limited function that is uniquely defined by its sample with this particular  rate.
 This
principle defines the choice of the sampling rate in almost all signal processing protocols.
A similar principle works for discrete time processes and defines  recoverability of sequences from their subsequences. For example, sequences
with a spectrum located inside the interval  $(-\pi/2,\pi/2)$ can be recovered from their
decimated subsequences consisting of all terms with even numbers allows to recover sequences.

 Numerous  extensions of the sampling theorem  were obtained,
 including
the case of nonuniform sampling and restoration of the signal with mixed samples;
see  \citet{BM,Cai,jerri,F95,La,La2,OU08,U1,U2,U50,V87,V01,Z} an literature therein. There were works studying
 possibilities to reduce  the set of sampling points
required for restoration  of the underlying functions.
In particular, it was found   that a band-limited function can be recovered without error from an oversampling sample sequence  if a finite number of sample values is unknown, or if an one-sided half of any oversampling sample sequence is unknown \citet{V87}.
It was found   \citet{F95}
that the function can be recovered without error  with a missing equidistant  infinite subsequence consistent of  $n$th
member of the original sample sequence, i.e. that each $n$th member is redundant, under some additional constraints on the
oversampling parameter. The constraints are such that the oversampling parameter is increasing if  $n\ge 2$ is decreasing. There is also an approach based on the so-called Landau's phenomenon \cite{La,La2};
see \cite{BM,La,La2,OU08,U1,U2} and a recent literature review in \cite{OU}.
This approach
allows  arbitrarily sparse discrete uniqueness sets in the time domain for a fixed spectrum range; the focus is on the uniqueness problem rather than on algorithms for recovery.
Originally, it was shown in  \cite{La} that the set of sampling points representing small deviations of integers is  an uniqueness set for classes of functions with an arbitrarily large measure of the spectrum range, given that there are periodic gaps in the spectrum and that the spectrum range is bounded. The implied sampling points were not equidistant and have to be calculated.
This result was significantly extended. In particularly,    similar uniqueness for functions with unbounded spectrum range and for  sampling points
 allowing a simple explicit form  was established in \cite{OU08} (Theorem 3.1 therein). Some generalization  for multidimensional case were obtained in \cite{U1,U2}.
However, as was emphasized in \cite{La2},  the  uniqueness theorems  do not ensure  robust  data recovery; moreover, it is also known that any sampling below the Nyquist rate  cannot be stable in this sense \cite{La2,U2}.

 The present paper readdresses the problem of  sparse
 sampling using a different approach. Instead of analyzing continuous time functions with spectrum degeneracy,  it focuses  on analysis of spectrum characteristics of  sequences
that can be recovered from their periodic subsequences,  independently from the sampling problem for the continuous time functions.
  The goal was to describe for a given integer $m>0$, class of sequences $\ww x\in\ell_2$,
featuring the following properties:
 \begin{itemize}
\item[(i)]  $\ww x$ can be recovered from a subsample $\ww x(km)$;
\item[(ii)] these processes are everywhere dense in $\ell_2$, i.e. they can approximate  any  $x\in\ell_2$.
\end{itemize}
\par

We found
a solution based on a special kind of
spectrum degeneracy. This degeneracy does not assume that there are spectrum gaps for a frequency characteristic such as Z-transform.

Let us describe briefly introduced below classes of processes with these properties.  For a process
 $x\in \ell_2$, we consider an auxiliary  "branching" process, or a set of processes  $\{\w x_d\}_{d=-m+1}^{m-1}\in \ell_2$
 approximating alterations of the original process.
It appears that certain conditions of periodic degeneracy of the spectrums for all $\w x_d$,
  ensures that it is possible to compute a new representative sequence  $\ww x$ such that there exists a  procedure for recovery $\ww x$ from the subsample $\{\ww x(km)\}$. The procedure
  uses  a  linear predictor  representing a modification of the predictor from \cite{D12a} (see the proof of  Lemma \ref{ThOdd} below). Therefore,  desired properties (i)-(ii) hold.
  We interpret this  as  a spectrum degeneracy of a new kind for $\ww x$  (Definition \ref{defAD1} and Theorems \ref{ThDeg}--\ref{ThD2} below).
In addition, we show  that the procedure of recovery of any finite part of the
  sample  $\ww x(k)$ from the subsample $\ww x(km)$ is robust with respect to noise contamination and data truncation, i.e. it
  is  a well-posed problem in a certain sense (Theorem \ref{ThA}).

Further, we apply the results sequences to  the
sampling of  continuous time band-limited functions.  We  consider  continuous time band-limited functions which samples are sequences
  $\ww x$ featuring the degeneracy  sequences mentioned above. As a corollary, we found that  these functions are uniquely defined by $m$-periodic subsamples
   of their equidistant  samples with  the critical Nyquist rate  (Corollary \ref{corr1}  and Corollary \ref{ThAC}).
 This allows to bypass, in a certain sense, the restriction on the sampling rate described by the critical   Nyquist rate, in the sprit of \cite{BM,La,La2,OU08,U1,U2}.
 The difference is that  we consider equidistant sparse sampling; the sampling points in \cite{BM,La,La2,OU08,U1,U2} are not equidistant, and this was essential. Further, our method is based on a recovery
 algorithm for sequences;  the results in \cite{BM,La,La2,OU08,U1,U2} are for the uniqueness problem and do not cover recovery algorithms.
 In addition, we consider different classes of functions; Corollary \ref{corr1}  and Corollary \ref{ThAC}
 cover band-limited functions, on the other hands, the uniqueness result
 \cite{OU08} covers function with unbounded spectrum periodic gaps. These gaps are not required
 in Corollary \ref{corr1}; instead,we request that sample series for the underlying function featuring spectrum degeneracy according to Definition \ref{defAD1} below.}

\subsubsection*{Some definitions and notations}
We denote by $L_2(D)$ the usual Hilbert space of complex valued
square integrable functions $x:D\to\C$, where $D$ is a domain.

We denote by $\ZZ$  the set of all integers. Let $\ZZ_q^+=\{k\in\ZZ:\ k\ge q\}$, let $\ZZ_q^-=\{k\in\ZZ:\ k\le q\}$, and let $\ZZ_{[a,b]}=\{k\in\ZZ:\ a\le a\le b\}$.

\par
For a set $G\subset \ZZ$ and $r\in[1,\infty]$,
we denote by $\ell_r(G)$ a Banach
space of complex valued sequences $\{x(t)\}_{t\in G}$
such that
$\|x\|_{\ell_r(G)}\defi \left(\sum_{t\in G}|x(t)|^r\right)^{1/r}<+\infty$ for $r\in[1,+\infty)$,
and $\|x\|_{q(G)}\defi \sup_{t\in G}|x(t)|<+\infty$ for $r=\infty$.
We denote  $\ell_r=\ell_r(\ZZ)$.

We denote by $B_r(\ell_2)$ the closed ball of radius $r>0$ in $\ell_2$.

\par
Let $D^c\defi\{z\in\C: |z|> 1\}$, and let $\T\defi \{z\in\C: |z|=1\}$.
\par
For  $x\in \ell_2$, we denote by $X=\Z x$ the
Z-transform  \baaa X(z)=\sum_{k=-\infty}^{\infty}x(k)z^{-k},\quad
z\in\C. \eaaa Respectively, the inverse Z-transform  $x=\Z^{-1}X$ is
defined as \baaa x(k)=\frac{1}{2\pi}\int_{-\pi}^\pi
X\left(e^{i\o}\right) e^{i\o k}d\o, \quad k=0,\pm 1,\pm 2,....\eaaa
For $x\in \ell_2$, the trace $X|_\T$ is defined as an element of
$L_2(\T)$.

 Let  $\HHH^r(D^c)$ be the Hardy space of functions that are holomorphic on
$D^c$ including the point at infinity  with finite norm
\baaa
\|h\|_{\HHH^r(D^c)}=\sup_{\rho>1}\|h(\rho e^{i\o})\|_{L_r(-\pi,\pi)},\quad r\in[1,+\infty].
\eaaa

\index{Note that Z-transform defines a bijection
between the sequences from $\ell_2^+$ and the restrictions (i.e.,
traces) $X|_{\T}$ of the functions from $\HHH^2(D^c)$ such that  $\overline{X\ew} =X\left(e^{-i\o}\right)$ for $\o\in\R$; see, e.g., \cite{Linq}, Section 4.3.
If $X\ew\in L_1(-\pi,\pi)$ and $\overline{X\ew} =X\left(e^{-i\o}\right)$, then $x=\Z^{-1}X$
is defined as an element of $\ell_\infty^+$.}

In this papers, we focus on processes without spectrum gaps
of positive measure on $\T$ but with a spectrum gap at isolated points on $\T$ where Z-transforms vanishes with a certain rate.

Let $L>1$ be given.
\par
For  $\b\in (-\pi,\pi]$,  $q>1$, $c>0$, and $\o\in(-\pi,\pi]$, set \baaa
\WW(\o,\b,q,c)=\frac{1}{L}\max\left(L,\exp\frac{c}{|e^{i\o}-e^{i\b}|^{q}}\right).
\label{hdef}\eaaa


For $r>0$, let $\X_{\b}(q,c,r)$ be the set of all   $x\in \ell_2$ such that
\baa \int_{-\pi}^\pi |X\ew|^2 \WW(\o,\b,q,c)^2d\o\le r,
\label{hfin}\eaa\ where $X=\Z x$.

\noxxx{For $m\in \ZZ_1^+$ and $s\in\ZZ$, let $\SS_{m,s}:\ell_2\to\ell_2$ be a mapping representing
decimation of a sequence such that
 $y(t)=x(t)\Ind_{\{(t+s)/m\in\ZZ\}}$ for $y=\SS_{m,s}x$.}

 Let $\SSS_{m,s}:\ell_2\to\ell_2$ be a mapping representing
extraction of a subsequence such that $\w y(k)=x(km-s)$ for all $k\in\ZZ$ for $\w y=\SSS_{m,s}x$.

\def\wh{h}\def\wH{H}
Let $\w\K^+$ be the class of functions $\wh:\ZZ\to\R$
such that $\wh (t)=0$ for $t<0$ and such that $\wH=\Z\wh\in H^\infty(D^c)$. Let $\w\K^-$ be the class of functions $\wh:\ZZ\to\R$
such that $\wh (t)=0$ for $t>0$ and such that $\wH=\Z\wh\in H^\infty(D)$. Let $\w\K$ be the linear span of $\w\K^+\cup\w\K^-$.

\section{Preliminary  results: predictability of sequences}
\label{SecM}
\def\RR{{\cal R}}
Up to the end of this paper, we assume that we are given $m\ge 1$, $m\in\ZZ$.

\begin{definition}\label{defP}
 Let $\X\subset \ell_2$ be a class of
processes.
 We say that the class $\X$ is  uniformly $\ell_2$-predictable on finite horizon   if for any $\e>0$ and any integers $n>0$ (any $n <0$) there exists $\wh\in \w\K^{+}$ (respectively, $\wh\in \w\K^{-}$ for $n<0$) and $\psi\in\ell_\infty$  such that $\inf_k|\psi(k)|>0$, $\sup_k|\psi(k)|<+\infty$, and   \baaa \sum_{t\in\ZZ}|x(t+n)- \w x(t)|^2\le \e\quad
\forall x\in\X, \label{predu}\eaaa where \baaa
\w x(t)\defi
\psi(t)^{-1}\sum_{s\in\ZZ}\wh(t-s)\psi(s)x(s).
\eaaa
\end{definition}

In Definition \ref{defP}, the use of $\wh\in\w \K^-$ means causal extrapolation, i.e. estimation  of $\w x(t+n)$
for $n>0$ is based on observations $\{x(k)\}_{k\le t}$, and the use of $\wh\in\w \K^+$ means anti-causal extrapolation, i.e. estimation  of $\w x(t+n)$ for $n<0$
is based on observations  $\{x(k)\}_{k\ge t}$.

\begin{lemma}\label{ThV}
For any $\b\in(-\pi,\pi]$, $\qq>0$, $c>0$,    and $r>0$, the class of sequences
$\X_{\b}(\qq,c,r)$ is uniformly predictable on finite horizon in the sense of Definition
\ref{defP}  with $\psi(t)=e^{i\t t}$ and $\t=\b-\pi$.
\end{lemma}
\noxxx{\begin{remark} {\rm As is seen from the proof below, Lemma \ref{ThV} is still valid if $\w\K$
in  Definition \ref{defP} is replaced by a smaller set $\w\K^-\cup \w\K^+$.}
\end{remark}}
\subsubsection*{Robustness of prediction}
\begin{definition}\label{defRob}  Let $\X\subset \ell_2$ be a set of sequences.
Consider  a problem of predicting  $\{x(k)\}_{\in\ZZ_{[-M,M]}}$  using observed
 noise contaminated sequences $x=\ww x+\xi$, where  $\ZZ_{[-M,M]}=\{k\in\ZZ,\ |k|\le M\}$,
 $\ww x\in\X$, and where $\xi\in\ell_2$ represents a noise. Suppose that only
truncated traces of observations of
$\{x(k)\}_{k\in\ZZ_{[-N,-M-1]}}$ available (or only  traces $\{x(k)\}_{k\in\ZZ_{[M+1,N]}}$ are available), where  $N>M$  is an integer. We say that the class $\X$ allows uniform and robust prediction if,  for   any integer $M>0$ and any $\e>0$,  there exists $\rho >0$,  $N_0>M$,
a  set of sequences $\{\psi_{t}(\cdot)\}_{t\in\ZZ_{[-M,M]}}$ such that $\psi_t\in\ell_\infty$, $\inf_k|\psi_t(k)|>0$, $\sup_k|\psi_t(k)|<+\infty$ for any $t$, and a set
 $\{\wh_{t}(\cdot)\}_{t\in\ZZ_{[-M,M]}}\subset \K^+$ (or a set
 $\{\wh_{t}(\cdot)\}_{t\in\ZZ_{[-M,M]}}\subset \K^-$ respectively) such that \baaa
\max_{t\in\ZZ_{[-M,M]}}| x(t)-\w x(t)|\le \e \quad
\forall \ww x\in\X_{\b}(\qq,c,r),\quad\forall  \xi\in B_\rho(\ell_2), \label{predUU}\eaaa
  for any  $N>N_0$ and for  \baaa
\w x(t)=
\psi_t(t)^{-1}\sum_{s\in \TT_0,\ M<|s|\le N}\wh_{t}(t-s)\psi_t(s)x(s),\quad t\in\ZZ_{[-M,M]},\eaaa
 with $\psi_t(s)=e^{i\t s}$ and $\t=\b-\pi$.
  \end{definition}
The following theorem shows that predicting  is robust with respect to noise contamination and truncation.
\begin{lemma}\label{ThR}  For given $\b\in(-\pi,\pi],\qq>0,c>0,r>0$, the class  $\ww x\in\X_{\b}(\qq,c,r)$ allows uniform and robust prediction  in the sense of Definition \ref{defRob}  with $\psi(t)=e^{i\t t}$ and $\t=\b-\pi$.   \end{lemma}

\par
\section{The main results: recoverability of sequences  from subsequences}
We consider  some problems of recovering sequences from their decimated subsequences.

\begin{definition}\label{defR}
Let $\X\subset \ell_2$ be a class of sequences. Let $\TT\subset \ZZ$.
We say that the class $\X$ is  uniformly recoverable  from observations of $x$ on  $\TT$  if, for any integer $M>0$ and any $\e>0$, there exists
a  set of sequences $\{\psi_{t}(\cdot)\}_{t\in\ZZ_{[-M,M]}}$ such that $\psi_t\in\ell_\infty$, $\inf_k|\psi_t(k)|>0$, $\sup_k|\psi_t(k)|<+\infty$ for any $t$, and
a set $\{\wh_{t}(\cdot)\}_{t\in\ZZ_{[-M,M]}} \subset \w\K$ such that \baaa
\max_{t\in\ZZ_{[-M,M]}} |x(t)- \w x(t)|\le \e\quad
\forall x\in\X, \label{predU}\eaaa where  \baaa
\w x(t)=\psi_t(t)
\sum_{s\in \TT}\wh_{t}(t-s)\psi_t(s)x(s),\quad t\in\ZZ_{[-M,M]}.\eaaa
\end{definition}

\begin{definition}\label{defDeg}
Let   $\qq>0$, $r>0$,   and $r>0$ be given.
 We say that $x\in \ell_2$
 features $m$-braided \index{tangled} spectrum degeneracy    if, for $d=-m+1,...,m-1$,
 there exist  numbers $\b_d\in (-\pi,\pi]$  and  $\yy _d\in
 \X_{\b_d}(\qq,c,r)$
 such that the following  holds:
 \begin{enumerate}
\item  All $\b_d$ are different  for $d=-m+1,....,m-1$.
\item
\baa
 &&\yy _d(k)=\yy _0(k),\quad k\le 0,\quad d>0,\nonumber\\
 && \yy _d(k)=\yy _0(k),\quad k\ge 0,\quad d<0.
\label{yPm} \eaa
\item
  \baa
 &&x(k)=\sum_{d=0}^{m-1}\yy _d\left(\frac{k+d}{m}\right)\Ind_{\left\{ \frac{k+d}{m}\in\ZZ\right\}},\quad k\ge 0,\nonumber\\
 &&x(k)=\sum_{d=-m+1}^{0}\yy _d\left(\frac{k+d}{m}\right)\Ind_{\left\{ \frac{k+d}{m}\in\ZZ\right\}},\quad k\le 0.
\label{xPm} \eaa
\end{enumerate}
We denote by $\oo\b$ the corresponding set $\{\b_d\}_{d=-m+1}^{m-1}$, and we denote by ${\cal B}$
the set of all $\oo\b$ with the features described above. We denote by $\PP_{m,\oo\b}(\qq,c,r)$ the set  of  all  sequences  $x$ featuring this degeneracy, and we denote
 \baaa
 \PP_{m,\oo\b}=\cup_{\qq>1,c>0,r>0}\PP_{m,\oo\b}(\qq,c,r),\quad \PP_m=\cup_{\oo\b\in{\cal B}}\PP_{m,\oo\b}.
 \eaaa
  \end{definition}

\begin{theorem}\label{ThDeg} For any $\oo\b\in{\cal B}$, $q>1$, $c>0$, $r>0$, the class  $\PP_{m,\oo\b}(\qq,c,r)$ is uniformly recoverable from observations on the set $\{k\in \ZZ:\ k/m\in\ZZ,\  |k|>s\}$
in the sense of Definition \ref{defR}.
\end{theorem}

\begin{corollary}\label{corrU}
 A sequence   $x\in\PP_m$ is uniquely defined by its subsequence $\{x(km)\}_{k\in \ZZ}$. Moreover, for any $s>0$, a sequence   $x\in\PP_m$ is uniquely defined by its subsequence $\{x(km)\}_{k\in \ZZ,\ |k|>s}$.

\end{corollary}


\begin{theorem}\label{ThDense}
For any $\oo\b\in{\cal B}$, any $x\in \ell_2$, and any $\e>0$, there exists  $\w x\in \PP_{m,\oo\b}$  such that \baa
\|x-\w x\|_{\ell_2}\le \e.
\label{xdd}\eaa
In particular, the  set $\PP_{m}$ is everywhere dense in $\ell_2$.
\end{theorem}

Let us compare these results with the result of \cite{Ser}, where a method was suggested
 for recovery from observations of subsequences
of missing values  oversampling sequences
for band-limited continuous time  functions.  The  method \cite{Ser} is also applicable
for general type band-limited sequences from $\ell_2$.
The algorithm in  \cite{Ser} requires to observe quite large number of subsequences, and this number is increasing as the size of spectrum gap  on $\T$ is decreasing.
Theorems \ref{ThDeg} and \ref{ThDense} ensures recoverability with   just one subsequence for a class of sequences that everywhere dense in $\ell_2$.

\subsubsection*{Robustness of recovery}

The following theorem shows that recovery of a finite part of a sequence from $\PP_m$
from its $m$-periodic subsequence is robust with respect to noise contamination and truncation.

\begin{theorem}\label{ThRR} Let $m\in\ZZ_1^+,\oo\b\in{\cal B},\qq>1,c>0,r>0$  be given.
Consider  a problem of recovery of  the set  $\{x(k)\}_{k=-M}^M$   from observed subsequences of
 noise contaminated sequences $x=\ww x+\xi$, where  $M\in\ZZ_0^+$,  $\ww x\in\PP_{m,\oo\b}(\qq,c,r)$ and where $\xi\in\ell_2$ represents a noise. Suppose that only
truncated traces of observations of
$\{x(k),\ k\in\TT_0,\ M<|k|\le N\}$ are available, where  $N>0$  is an integer.
Then for   any integer $M>0$ and any $\e>0$,  there exists $\rho >0$,  $N_0>0$,
a  set of sequences $\{\psi_{t}(\cdot)\}_{t\in\ZZ_{[-M,M]}}$ such that $\psi_t\in\ell_\infty$, $\inf_k|\psi_t(k)|>0$, $\sup_k|\psi_t(k)|<+\infty$ for any $t$, and a set
 $\{\wh_{t}(\cdot)\}_{t\in\ZZ_{[-M,M]}}\subset \w\K$ such that \baaa
\max_{t\in\ZZ_{[-M,M]}}| x(t)-\w x(t)|\le \e \quad
\forall \ww x\in\PP_{m,\oo\b}(\qq,c,r),\quad\forall  \xi\in B_\rho(\ell_2), \label{robRR}\eaaa
  for any  $N>N_0$ and for  \baaa
\w x(t)=\psi_t(t)
\sum_{s\in \TT_0,\ M<|s|\le N}\wh_{t}(t-s)\psi_t(s)x(s),\quad t\in\ZZ_{[-M,M]}.\eaaa
 \end{theorem}
\noxxx{
\section{Applications to sampling in continuous time}

Up to the end of this paper, we assume that we are given  $\O>0$ and  $\tau>0$  such that $\O\le\pi/\tau$.
 We will denote $t_k=\tau k$, $k\in\ZZ$.

The classical
Nyquist-Shannon-Kotelnikov Theorem states that {
a band-limited function
$f\in \BLO$ is uniquely defined by the  sequence $\{f(t_k)\}_{k\in \ZZ}$.}

The sampling rate $\tau=\pi/\O$ is called the critical Nyquist rate for $f\in \BLO$.
If $\tau<\pi/\O$, then, for any finite set $S$ or for $S=\ZZ^\pm$, any
$f\in\BLO$ is uniquely defined by the values   $\{f(t_k)\}_{k\in\ZZ\backslash S}$, where  $t_k=\tau k$, $k\in\ZZ$; this
was established in \citet{F91,V87}.   We cannot claim the same for some infinite sets of missing values.
For example, it may happen that
$f\in \BLO$  is not uniquely defined by the values $\{f(t_{mk})\}_{k\in\ZZ}$  if the sampling rate for this  sample
 $\{f(t_{mk})\}_{k\in\ZZ}$ is lower than is the so-called critical Nyquist rate implied by the Nyquist-Shannon-Kotelnikov Theorem; see more examples in \citet{F95}. We address this problem below.

\begin{theorem}\label{ThCTD}  For any  $\d>0$ and any $\ww x\in \PP_{m,\oo\b}(\qq,c,r)$, there exists
a unique $\ww f\in \BLOO$ such that
$\ww f(t_k)=\ww x(k)$. This
$\ww f$  is uniquely defined by the sampling subsequence  $\{\ww f(t_{mk})\}_{k\in \ZZ}$.
\end{theorem}
\par
 We denote by $\F_{m,\oo\b}(\qq,c,r)$ the set of  all functions $\ww f\in \BLOO$ constructed for all  $\ww x\in \PP_{m,\oo\b}(\qq,c,r)$
as described in Theorem \ref{ThCTD}. We denote
 \baaa
 \F_{m,\oo\b}=\cup_{\qq>1,c>0,r>0}\F_{m,\oo\b}(\qq,c,r),\quad \F_m=\cup_{\oo\b\in{\cal B}}\PP_{m,\oo\b}.
 \eaaa

\begin{remark}
Theorem  \ref{ThCTD} implies  that $\F_m$ can be considered as a class of functions  featuring  spectrum degeneracy of a new kind. The sequences $\ww f$ from this class can be recovered from  a sample taken below the Nyquist rate, i.e. from the sample $\{\ww f(t_{mk})\}_{k\in \ZZ}$, where $t_k=\tau k$, and
$\tau>0$ is  such that $\O\le\pi/\tau$.
\end{remark}

\begin{corollary}\label{corr1} Let $\e>0$, $\oo\b\in{\cal B}$, and $f\in \BLOO$ be given, and  let $x\in\ell_2$ be defined as $x(k)=f(t_k)$
for $k\in \ZZ$. Then there exists $\w\e>0$   such that any    $\ww x\in\PP_{m,\oo\b}$ satisfying
\baaa
\|x-\ww x\|_{\ell_2}\le \w\e
\label{xd1}\eaaa
is such that  $\ww f\in \BLOO$  defined as described in Theorem \ref{ThCTD}  is uniquely defined by the values   $\{\ww f(t_{mk})\}_{k\in \ZZ}$, and is such that
\baaa
\|f-\ww f\|_{L_p(\R)}\le \e,\quad p=2,+\infty.
\eaaa
\end{corollary}
\begin{remark}{\rm
In Corollary  \ref{corr1}, $\ww f$ can be viewed as a result of an arbitrarily small adjustment of $f$.
This $\ww f$ is uniquely defined by sample  with a
sampling distance  $m \tau$, where   $\tau$ is a distance smaller than the distance defined by  critical Nyquist rate for $f$.
Since the value $\e$ can be arbitrarily small, and $m$ and $M$ can be arbitrarily large in Theorem \ref{corr1} and Corollary \ref{ThRR}, one can say  that the restriction on the sampling rate defined by the  Nyquist rate is bypassed, in a certain sense.
}\end{remark}

 \begin{remark}{\rm
Theorem \ref{ThCTD} and Corollary \ref{corr1} represent uniqueness results for equidistant uniqueness set of functions $\ww f$ from
a special subset of the class of band-limited functions  in the sprit of \cite{La,La2,OU08,OU,U1,U2}.
 These works considered more general function  with restrictions on the spectrum domain, including unbounded domains;  the uniqueness sets of times in these works are not equidistant.
 Corollary \ref{corr1} considers band-limited functions only. However, this Corollary allows  and  equidistant uniqueness sampling sets $\{t_{mk}\}_{k\le 0,\ k\in \ZZ}=\{m\tau k\}_{k\le 0,k\in\ZZ}$.
The corresponding functions  $\ww f$  are described as extrapolants of  sampling sequences from a special class rather then defined by the spectrum; this is also different from   \cite{La,La2,OU08,OU,U1,U2}.

}\end{remark}
\begin{corollary}\label{ThRRC} Under the assumptions and notations of Corollary \ref{corr1},
consider  a problem of recovery of the set $\{\ww f(t_n)\}_{n=-M}^M$
from a noise contaminated truncated series  of observations  $\{\ww f(t_{mk})+\xi(mk)\}_{k=-N}^N$, where $M>0$ and $N>0$ are integers,  $f\in\F_{m,\oo\b}(\qq,c,r)$, and where  $\xi\in\ell_2$ represents a noise contaminating the observations. Then
 this problem
 is well-posed in the following sense: for any $M$ and any  $\ww\e>0$,
 there exists a recovery algorithm and $\d>0$ such that there exists an integer $N_0>0$ such the recovery error is such that
 \baaa
 \max_{n=-M,..,M}|\ww f(t_n)-\ww f_E(t_n)|\le \ww\e \quad \forall f\in\F_{m,\oo\b}(\qq,c,r),\quad \forall\xi\in B_\d(\ell_2),\quad \forall N\ge N_0.
 \eaaa
Here $\ww f_E(t_n)$ is the estimate of $\ww f(t_n)$ obtained by the corresponding recovery algorithm.
\end{corollary}
\begin{remark}{\rm \comm{ADDED IN 2018} Corollary \ref{ThRRC} establishes
robust recoverability only for a finite sample with $2M$
members; this does not satisfy criteria of the stable sampling in the sense of  \cite{La2}.  Therefore, this result does not overcome restrictions imposed  by the so-called Landau's rate \cite{La2}  for the rate of stable sampling for continuous time functions.
}\end{remark}

\subsubsection*{The case of  non-bandlimited continuous time functions}
Technically speaking, the classical sampling theorem is applicable to band-limited continuous time  functions only.
However, its  important role in signal processing is based on applicability  to more general functions since they allow  band-limited approximations: any $f\in L_1(\R)\cap L_2(\R)$ can be approximated by bandlimited functions
$f_\O\defi \F^{-1}(F\Ind_{[-\O,\O]})$ with $\O\to +\infty$, where $F=\F f$, and where $\Ind$ is the indicator function. This still means that the sampling frequency has to be increased along with $\O$:  for a given $\tau>0$, the sample  $\{f_\O(\tau k)\}_{k\in\ZZ}$
defines the function $f_\O$ if  $\tau\le \pi/\O$. A related problem is aliasing of continuous time processes after time discretization.
Corollary \ref{corr1}  allows to overcome this obstacle in a certain sense as the following.

\begin{corollary}\label{corr2}   For any $f\in L_1(\R)\cap L_2(\R)$,  $\e>0$, and $\Delta>0$, there  exists $\oo\O>0$   such that the following holds for any   $\O\ge \oo\O$:
 \begin{enumerate}
 \item
 $\sup_{t\in\R} |f(t)-f_{\O}(t)|\le\e$, where  $f_{\O}= \F^{-1}F_{\O}$, $F_{\O}=F\Ind_{ [-\O,\O]}$, $F=\F f$.
\item
The function $f_{\O}$ belongs to $\BLO$, and, for $\tau=\pi/\O$, $t_k=\tau k$,   satisfies the assumptions of Corollary \ref{corr1}. For this function, for any $\e>0$ there  exists $\ww f\in \BLO$ such that
 $\sup_{t\in\R} |f_{\O}(t)-\ww f(t)|\le\e$ and that $\ww f$  is uniquely defined by the values   $\{\ww f(\t_{k})\}_{k\in \ZZ}$ for an equidistant sequence of sampling points $\t_k =k \Delta$, $k\in\ZZ$, for any $s\in\ZZ$, with robustness described in Corollary \ref{ThRRC}.
\end{enumerate}
\end{corollary}
}
\section{Proofs}
It suffices to consider prediction on $n$ steps forward
with $n>0$. The case where $n<0$ can be considered similarly.

\subsubsection*{Special predicting kernels}

The proofs are based on special predicting kernels  and their transfer functions representing modification
of kernels and transfer functions  introduced in  \cite{D12a}.

Let us introduce transfer functions and its inverse Z-transform
   \baa
 \w H_n(z)=\w H_{n,\dm}(z) \defi z^n V(z^{\dm })^n,\quad z\in\C,\qquad \w h_n=\Z^{-1}\w H_n, \label{wH}\eaa
where
 \baaa
 V(z)\defi 1-\exp\left[-\frac{\g}{z+ 1-\g^{-\w r}}\right],
\label{wK}
\eaaa
and where $\w r>0$ and $\g>0$ are  parameters. \index{Let $r > 2/(q-1)$ for some $q>1$.}
 Functions $V$ were introduced in  \cite{D12a}.
 In the notations from \cite{D12a}, $\w r=2\mu/(q-1)$, where $\mu>1$ and $q>1$ are parameters; $q$ describes  the required  rate of spectrum degeneracy.
We assume that $r$ is fixed, and we consider variable $\g\to +\infty$.

In the proof below, we will show that $\w H_n\ew$ approximates $e^{in\o}$ and therefore defines
a linear $n$-step predictor with the kernel $\w h_n=\Z^{-1}\w H_n$.

 The predicting kernels $\w h_n$ are real valued, since $\w H_n\,\!\left(\oo z\right)=\overline{\w H_n\,\!\left(z\right)}$.

\xxxonly{
{\em Proof of Lemma \ref{ThV}}.
Lemma \ref{ThV} represents a special case of Theorem 2.2 from \cite{D18x}, where $m=\nu=1$,
in the notations   of \cite{D18x}. $\Box$

{\em Proof of Lemma \ref{ThR}}.
Lemma \ref{ThR} represents a special case of Theorem 2.4  from \cite{D18x}, where $m=\nu=1$,
in the notations   of \cite{D18x}. $\Box$
}

\noxxx{
{\em Proof of Lemma \ref{ThV}}.
\subsubsection*{The case where $\b=\pi$ }
\begin{lemma}\label{lemmaX} Lemma \ref{ThV} holds for the case where $\b=\pi$. \end{lemma}

{\em Proof of Lemma \ref{lemmaX}}.
The proof represents a modification of the proof from
\cite{D12a}, where the case of $n=1$ was considered and where
the spectrum degeneracy were assumed to be at a single point only. In addition,  represents a modification of the proof from \cite{D16x}, where the case of $\b=\pi$ was considered.

Consider an arbitrary  $x\in\X_{\pi}(\qq,c,r)$. We denote  $X\defi \Z x$ and
$\w X\defi \w H_n\,\! X$.

\def\OO{W}
Let $\a=\a(\g)\defi 1-\g^{-r}$. Clearly, $\a=\a(\g)\to 1$  as $\g\to+\infty$.

Let $\OO(\a)=\arccos(-\a)$, let $D_+(\a)=(-\OO(\a),\OO(\a))$, and let
 $D(\a)\defi[-\pi,\pi]\backslash D_+(\a)$. We have that  $\cos(\OO(\a))+\a=0$, $\cos(\o)+\a>0$ for  $\o\in D_+(\a)$,
and $\cos(\o)+\a<0$ for  $\o\in D(\a)$.
\par

It was shown in \citet{D12a} that the following holds:
\begin{itemize}
\item[(i)] $V(z)\in \HHH^{\infty}(D^c)$ and $zV(z)\in
\HHH^{\infty}(D^c)$.
\item[(ii)] $V(e^{i\o})\to 1$ for all  $\o\in (-\pi,\pi)$ as  $\g\to +\infty$.
\item[(iii)] If $\o\in D_+(\a)$  then $\Re \left(\frac{\g}{e^{i\o}+\a}\right) >0$ and $|V\ew -1|\le 1$.
    \item[(iv)] For any $c>0$,
there exists $\g_0>0$ such that for any $\g\ge \g_0$ and for $V$
selected with $\a=\a(\g)$  we have
\baaa
\int_{D(\a)}|V\ew-1|^2
\WW(\o,\pi,q,c)^{-2}d\o\le 2\arccos(\a).
\eaaa
\end{itemize}

Without a loss of generality, we assume below that $\g>\g_0$.

From the properties of $V$, it follows that the following holds.
\begin{itemize}
\item[(i)] $V(z^{\dm})^n\in \HHH^{\infty}(D^c)$ and $ \w H_n(z)= z^nV(z^{\dm})^n\in
\HHH^{\infty}(D^c)$.
\item[(ii)] $V\left(e^{i \o \dm }\right)\to 1$  and $\w H_n\ew\to e^{in\o}$ for all  $\o\in (-\pi,\pi]\setminus R_{\dm}$ as  $\g\to +\infty$.
\item[(iii)]
\baaa
\Re \left(\frac{\g}{e^{i \o \dm }+\a}\right) >0,\quad
\left|V\left(e^{i\o \dm }\right) -1\right|\le 1, \quad\brea \o\in D_+(\a).
\eaaa
\end{itemize}
\noxxx{Figure ref{fig2} shows an example of the shape of error curves for approximation of
 the  forward one step shift operator. More precisely, they show
 the shape of $|\w H\ew -e^{i\o n}|$  for the transfer function   (\ref{wH}) with $n=4$,
  and  the shape of the corresponding predicting kernel $\w h$ with some selected  parameters.
}

\def\yo{x_n}
 \par Let $\yo (t)\defi x(t+n)$ and $X_n=\Z\yo$. In this case, $X_n(z)=z^n X(z)$.

  We have that
 \baaa
 \|X_n\ew-\w X\ew\|_{L_1(-\pi,\pi)}=I_1+I_2,
 \eaaa
where \baaa &&I_1=\int_{D(\a)}| X_n\ew-\w X\ew| d\o,\qquad\breakk
I_2=\int_{D_+(\a)}|X_n\ew-\w X\ew| d\o. \eaaa

By the assumption, (\ref{hfin}) holds.
 Hence

\baaa I_1&=&\|\w
 X\ew-X_n\ew\|_{L_1 (D(\a))}=
\|(\w H_n\ew-e^{i\o n})X\|_{L_1 (D(\a))}\nonumber\\
&\le& \|(V \ew -1)
 \WW(\o,\pi,q,c)^{-1}\|_{L_2 (D(\a))} \|X\ew \WW(\o,\pi,q,c)\|_{L_2(-\pi,\pi)}\nonumber
\\&\le& (2\arccos(\a))^{1/2} \|X\ew \WW(\o,\pi,q,c)\|_{L_2(-\pi,\pi)}.
\label{4s}\eaaa

  The last inequality holds by the properties of $V$. It follows that $I_1\to 0$ as $\g\to +\infty$
 uniformly over $x\in\X_{\pi}(\qq,c,r)$.

Let us estimate $I_2$. We have that
\baaa I_2=\int_{D_+(\a)} | (e^{i\o n}-\w H_n\,\!\left(e^{i\o  }\right)) X\ew|  d\o\break\le
\int_{D_+(\a)} |e^{i\o n}(1-V\left(e^{i\o \dm }\right)^n) X\ew|  d\o\\\le
\|1-V\left(e^{i\o \dm }\right)^n\|_{L_2(D_+(\a))} \|X\ew\|_{L_2(-\pi,\pi)}.\eaaa
\par
Further,
$\Ind_{D_+(\a)}(\o) |1-V\left(e^{i\o\dm}\right)^n| \to 0$ a.e. as $\g\to
+\infty$. By the properties of $V$, \baaa
\Ind_{D_+(\a)}(\o)|1-V\left(e^{i\o \dm }\right)^n | \le 2^n, \quad \forall \g:\ \g>\g_0.\eaaa   From Lebesgue Dominance Theorem,
it follows that
$\|1-V\left(e^{i\o \dm }\right)^n\|_{L_2(D_+(\a))}\to 0$ as $\g\to+\infty$. It follows that
$I_1+I_2\to 0$ uniformly over $x\in\X_{\b}(q,c,r)$.  Hence
\baa
\sup_{k\in\ZZ}|x(k+n)-\w x(k)|\to 0\quad \hbox{as}\quad \g\to+\infty\quad\brea
\hbox{uniformly over}\quad x\in \X_{\mnu \b}(q,c,r),
\label{conv}\eaa
 where the process $\w x$ is the output of the linear predictor defined by the kernel $\w h_n=\Z^{-1}\w H_n$ as
 \baa
  \w x(k)\defi \sum^{k}_{p=-\infty}\w h_n(k-p)x(p).\label{predict}
\eaa
Hence the
predicting kernels $\w h_n=\Z^{-1}\w H_n\,\!$ constructed for $\g\to +\infty$
are such as required. This
 completes the proof of Lemma \ref{lemmaX}, i.e. the proof of  Lemma \ref{ThV} for the case where  $\b=\pi$.
\subsubsection*{The case where $\b\in(-\pi,\pi]$ }
We are now in the position to complete the proof of Theorem \ref{ThV} for an arbitrarily selected  $\b\in(-\pi,\pi]$
 \par
 {\em Proof of  Lemma \ref{ThV}}.  Let $\t\defi \b-\pi$ and \baaa
\psi(t)\defi e^{i\t t},\qquad x_\b(t)\defi \psi(t)x(t),\qquad t\in\ZZ.
\eaaa
Let $x\in \X_{\mnu \b}(q,c,r)$ for some $\b$, $q$,$c$,$r$. Let $X_\b=\Z x_\b$ and $X=\Z x$.
In this case,
\baaa
X_\b\ew=\sum_{k\in\ZZ}e^{-i\o k}x_\b(k)=\sum_{k\in\ZZ}e^{-i\o k}\psi(k)x(k)\\
=\sum_{k\in\ZZ}e^{-i\o k +i\t k}x(k)=X\left(e^{i\o-i\t}\right).
\eaaa
We have that
\baaa
X_\b\left(e^{i\pi}\right)=X\left(e^{i\pi-i\t}\right)=X\left(e^{i\b}\right)=0.
\eaaa
Hence $x_\b\in \X_{\mnu \pi}(q,c,r)$. As we have established above,  $x_\b$  is predictable in the sense of Definition \ref{defP}. Let $\w h_n$ be the corresponding
predicting kernel which existence  is required by Definition \ref{defP} for the case where $\psi(t)\equiv 1$ such that the approximation $\w x_\b(t)$ of $x_\b(t+n)$
 is given by
  \baaa
\w x_\b(t)=\sum_{k=-\infty}^t \w h_n(t-k)y_\b(k).
\eaaa
These kernels
 were
 constructed in the proof of Lemma \ref{ThV}  above for the special case   $\b=\pi$.
It follows that  \baaa
\w x(t)=\psi(t)^{-1}\sum_{k=-\infty}^t \w h_n(t-k)\psi(t) x_(k).
\eaaa
 This completes the proof of Lemma \ref{ThV}.
$\Box$
\iindex{\begin{proposition}\label{proph}
\begin{enumerate}
\item The predicting kernels $\w h_n$ are real valued,
\item The predicting kernels $\w h_n$  are quite sparse:
\baa
 \w h_n(k)=0\quad  \hbox{if either}\quad (k+n)\notin \ZZ\quad \hbox{or}\quad  k<mn-n.
 \label{hm}\eaa
 \end{enumerate}
\end{proposition}
\par
{\em Proof.} To prove (i), it suffices to observe that    $\w H_n\,\!\left(\oo z\right)=\overline{\w H_n\,\!\left(z\right)}$.
Let us prove (ii).
By the choice of $V$, it follows that $|V(z)|\to 0$ as $|z|\to +\infty$. Hence  $v(0)=0$ for $v=\Z^{-1}V$ and \baaa
&&V(z^\dm)\breakk =z^{-\dm}v(1)+z^{-2\dm}v(2)+z^{-3\dm}v(3)+.. .\eaaa
 Clearly, we have that
\baaa
V(z^\dm)^n=z^{-n\dm}w(1)+z^{-(n+1)\dm}w(2)+z^{-(n+2)\dm}w(3)+...,\eaaa
where $w=\Z^{-1}(W)$ for $W(z)=V(z)^n$.
Hence \baaa
&&\w H_n(z^\dm)=z^nV(z^\dm)^n\breakk=z^{n-n\dm}w(1)+z^{n-(n+1)\dm}w(2)+z^{n-(n+2)\dm}w(3)+...\\
&&= z^{n-n\dm}\w h(n\dm-n)+z^{n-(n+1)\dm}\w h((n+1)\dm-n)\breakk+z^{n-(n+2)\dm}\w h((n+2)\dm-n)+...
\eaaa
Hence (\ref{hm}) holds. This completes the proof of Proposition \ref{proph}. $\Box$
}
\par
{\em Proof of lemma \ref{ThR}.}
It suffices to show that the error for recovery a singe term $x(n)$ for a given integer $n>0$  from the sequence $\{x(k)\}_{k\in \ZZ,\, k\le 0}$ can be made arbitrarily small is a well-posed problem; the proof for a finite set of values  to recover is similar.

The case where  $n<0$ and $\{x(mk)\}_{k\in \ZZ,\, k\ge 0}$ are observable
can be considered similarly.

We assume that $N>n$.

Let us consider an input sequence $x\in\ell_2$
such that \baa
x=\ww x+\eta,\label{xeta}
\eaa
where    $\eta(k)= \xi(k)\Ind_{\{|k|\le N\}}-\ww x(k)\Ind_{\{|k|>N\}}$.
In this case, (\ref{xeta}) gives that
$x(k)=(\ww x(k)+\xi(k))\Ind_{\{|k|\le N\}}$.

Let  $X=\Z x$, and let ${\cal N}\defi \Z\eta$, and let $\s=\|{\cal N}\ew\|_{L_2(-\pi,\pi)}$; this parameter  represents the
intensity of the noise.

Let us assume first that $\b=\pi$.

Let  an  arbitrarily small $\e>0$ be given.
Assume that  the parameters $(\g,\w r)$ of $\w H_{n}$ in (\ref{wH})
are selected such that \baa
|\oo x(n)-\oo x(0)|\le \e/2 \quad \forall  \oo x\in \X_{\mnu \b}(\qq,c,r),
\label{yy}\eaa
for  $\oo x=\Z^{-1}(\w H_n\oo X)$ and $\oo X=\Z\oo x$.
The kernel $\w h_n$ produces an estimate $\oo x(0)$  of
based on observations of $\{\oo x(km)\}_{k\le 0}$.

 Let  $\ww x\defi\Z^{-1}(\w H_n\ww X)$, where $\ww X=\Z \ww x$. By (\ref{yy}), we have that
\baaa  |\ww x(0)-\ww x(n)|\le\e/2.
\label{eps}\eaaa

 Let  $\w x\defi\Z^{-1}(\w H X)$, where $X=\Z x$.  We have that
$\w x=\ww x+\Z^{-1}(\w H {\cal N})$ and
 \baaa
  |\w x (0)-x(n)|\le   |\ww x (0)-x(n)|+|\Z^{-1}(\w H {\cal N})(0)|\le \e/2+ E_\eta,
  \eaaa
where
\baaa
E_{\eta}\defi \frac{1}{2\pi}\|(\w H\ew-e^{i\o n}){\cal N}\ew|\|_{L_1(-\pi,\pi)}\le \s (\kappa+1),
\eaaa
and where
\baaa
\kappa\defi \sup_{\o\in[-\pi,\pi]}|\w H_n\ew|.\eaaa
\par
We have  that  $\s\to 0$ as $N\to +\infty$ and $\|\xi\|_{\ell_2(-N,N)}\to 0$. If $N$ is large enough and  $\|\xi\|_{\ell_2(-N,N)}$ is small enough  such that $\s (\kappa+1)<\e/2$, then
$|\w x(0)-   x(n) |\le\e$.
This completes the proof of Lemma \ref{ThR} for the case where $\b=\pi$.

The prove for the case where  $\b\neq \pi$ follows from the case where $\b=\pi$, since
the processes $x(t),\oo x(t),\ww x(t),\w x(t),\ww x(t),$  presented in the proof  with $\b=\pi$ will be simply
converted to  processes $e^{-i\t t}x(t),e^{-i\t t}\oo x(t),e^{-i\t t}\w x(t),e^{-i\t t}\ww x(t)$, for $\t=\b-\pi$, similarly to the proof of Lemma \ref{ThV}.
 $\Box$
\begin{remark}{\rm
By the properties of $\w H_n$, we have that $\kappa\to +\infty$  as $\g\to +\infty$.
 This implies that error (\ref{eps}) will be increasing if $\w\e\to 0$ for any given $\s >0$.
This means that, in practice, the predictor  should not  target too small a size of the
error, since in it impossible to ensure that $\s=0$ due inevitable data truncation.
 }\end{remark}
 }
\def\TTT{{\cal I}}
%

Let us introduce  mappings $\M_d:\ell_2\to \ell_2$ for $d\in\ZZ_{[-m+1,m-1]}=\{-m+1,...,m-1\}$
such that, for  $x\in\ell_2$,  the sequence $x_d=\M_d x$ is defined by insertion of $|d|$ new members equal to
$x(0)$    as the following:
\baaa
&&\hphantom{}\hbox{(i)}\,\,\,\, x_0=x; \hphantom{xxxxx}  \hphantom{x_d(k)=x(0),\quad k= 0,1,...,d,\quad k<0}\,\,\nonumber \\
&&\hphantom{}\hbox{(ii)\,\, for} \quad d>0:\quad \breakk \hphantom{xxxxx}  x_d(k)=x(k),\quad k<0, \nonumber\\
&&\hphantom{xxxxx} x_d(k)=x(0),\quad k= 0,1,...,d, \nonumber\\
&&\hphantom{xxxxx} x_d(k)=x(k-d),\quad k\ge d+1;\nonumber\\
&&\hphantom{}\hbox{(iii) \,\, for} \quad d<0:\quad \breakk\hphantom{xxxxx}  x_d(k)=x(k),\quad k>0, \nonumber\\
&&\hphantom{xxxxx} x_d(k)=x(0),\quad k= d,d+1,...,0, \nonumber\\
&&\hphantom{xxxxx} x_d(k)=x(k-d),\quad k\le d-1.\hspace{-1cm}\label{xd2}
\eaaa

\par

{\em Proof of Theorem \ref{ThDeg}}. Let $x\in \PP_{m,\oo\b}(\qq,c,r)$.
By the definitions, (\ref{xPm}) holds with $\yy _d=\SSS_{m,0}x_d$, where  $x_d=\M_d x$.

Let $\ww\Y_d$ be the set of all $\yy _d=\SSS_{m,0} x_d$, where  $x_d=\M_d x$, for all  $x\in \PP_{m,\oo\b}(\qq,c,r)$.

Let $\ww\Y(\d)$ be the set of all ordered sets $\{\yy _d\}_{d=-m+1}^{m-1}$ such that $\yy _d\in \ww\Y_d(\d)$.

By Lemma \ref{ThV}, the sets $\ww\Y_d(\d)$  are  uniformly predictable in the sense of Definition
 \ref{defP}. It follows that,  for any $s\in\ZZ^+$ and $d\ge 0$, the sets $\ww\Y_d(\d)$  are  uniformly recoverable  in the sense of Definition
 \ref{defR} from observations   of $\{\yy _d(k)\}$ on $\{k\in \ZZ:\ k\le -s\}$.  Clearly, time direction in  Definition
 \ref{defP} can be reversed and therefore it can be concluded that, for $d<0$ and any $s\in\ZZ$, the sets $\ww\Y_d(\d)$  are  uniformly recoverable  in the sense of Definition   \ref{defP} from observations of $\{\yy _d(k)\}$ on
$\{k\in \ZZ:\     k\ge s\}$.

By the definitions again, we have that $\yy _0(k)=x(km)$ for all $k\in\ZZ$. For  $d>0$,  we have that
$\yy _d(k)=x(km-d)$ for $k\ge 1$ and $\yy _d(k)=x(km)$ for $k\le 0$.
  For  $d<0$,  we have that $\yy _d(k)=x(km-d)$ for $k\le -1$ and $\yy _d(k)=x(km)$ for $k\ge 0$.

   Let us apply this to establish recoverability of  $x(t)$ for $t\in\ZZ_{[0,M]}$ from the observations 
 $\{x(mt)\}_{t<-s}$ for any $s>0$. Let $d\in\ZZ_{[0,m-1]}$ be such that $(t+d)/m\in\ZZ$, i.e.
 $t=km-d$ for some $k\in\ZZ$, then
 $x(t)=y_d(k)$. As we found above, $y_d(k)$ can be recovered with kernels $h_k\in\K^-$ and functions $\psi_k\in\ell_{\infty}$  based on observations of
 $\{y_d(t)\}_{t<-s/m}=\{y_0(t)\}_{t<-s/m}=\{x(tm)\}_{t<-s}$ such as described in Definition \ref{defR} 
 and such that
 \baaa
 |y_d(t)- \w y_d(t)|\le \e\quad
\forall x\in\X, \label{predUUU}\eaaa where  \baaa
\w y_d(t)=\psi_t(t)
\sum_{s<M/m,\ s\in\ZZ}h_{t}(t-s)\psi_t(s)y_d(s)\brea=\psi_t(t)
\sum_{s<M/m,\ s\in\ZZ}h_{t}(t-s)\psi_t(s)x(ms).\eaaa
Similar arguments  establish recoverability of  $x(t)$ for $t\in\ZZ_{[-M,-1]}$ from the observations
 $\{x(mt)\}_{t>s}$ for any $s>0$. Therefore,  $x$  is  uniformly recoverable  in the sense of Definition   \ref{defR} from the
  observations of $x(t)$ on $\{t:\ t/m\in\ZZ,\ |k|\ge s\}$ for any $s>0$. This  completes the proof of Theorem \ref{ThDeg}. $\Box$
\par
{\em Proof of Theorem \ref{ThDense}}.  In the proof below, we consider $d\in\ZZ_{[-m+1,m-1]}=\{-m+1,...,m-1\}$.
\def\YY{Y}
 Let $x\in\ell_2$ be arbitrarily selected. Let
$x_d\defi \M_d x$ and $\yy_d\defi \SSS_{m,0}x_d$.

\par
Let $\xi_d\defi \yy_0-\yy_d$ and $\Xi_d\defi\Z \xi_d$.

Since $\yy_d(k)=\yy_0(k)$ for $k\le 0$ and $d>0$, and $\yy_d(k)=\yy_0(k)$ for $k\ge  0$ and $d<0$, it follows that $\xi_d(k)=0$ for $k\le 0$ and $d>0$, and $\xi_d(k)=0$ for $k\ge 0$ and $d<0$. In addition, $\Xi_0\equiv 0$ and $\xi_0\equiv 0$.

Further, let $D=\{k\in\ZZ:\ |k|\le m-1,\ k\neq 0\}$, and let
\baaa A(\o)=\prod_{d=-m+1}^{m-1}\varrho(\o,\b_d,q,c)^{-1},\quad
\a_d(\o)=1-\varrho(\o,\b_d,q,c)^{-1}.
\eaaa

By the choice of functions $\varrho$, we have that for large enough $\g>0$ that
\baa a_p\ew\varrho(\o,\b_d,q,c)=0\quad\hbox{a.e.},\quad p\neq d.
\label{arho}
\eaa
In addition, we have that,  for any $(L,\w r,c)$,  for large enough $\g$,
\baa (a_d(\o)-1)\varrho(\o,\b_d,q,c)\equiv 1.
\label{arho1}\eaa
We assume that $\g,L,\w r,c$ are selected  such that (\ref{arho}) holds.

Let \baaa
\w \YY_0\ew\defi  \YY_0\ew A(\o)\brea +
\sum_{p\in D} \Xi_p\ew a_p(\o).
\eaaa We have that $\w x_0\in \YY_{\b}(q,c,r)$, where $\w x_0=\Z^{-1}\w \YY_{0}$.

Furthermore,  by the definitions,
\baaa
\w \YY_d\ew=\YY_d\ew+\w \YY_{0}\ew-\YY_0\ew=\YY_0\ew A(\o)+W_d\ew,
\eaaa
where
\baaa
W_d\ew=\YY_d\ew+
\sum_{p\in D} [\YY_0\ew-\YY_p\ew] a_d(\o) -\YY_0\ew.
 \eaaa

By the choice of $A\ew$ and by (\ref{arho1}), it follows  for all $d$ that
\baa\int_{-\pi}^\pi |\YY_0\ew A(\o)|^2 \WW(\o,\b_d,q,c)^2d\o\le r.
\label{XA}\eaa

Let us show that,  for large enough $\g$, \baa \int_{-\pi}^\pi |W_d\ew|^2 \WW(\o,\b_d,q,c)^2d\o\le r.
\label{W}\eaa
We have that
\baaa
W_d\ew=\YY_d\ew-\YY_0\ew+
\sum_{p\in D} [\YY_0\ew-\YY_p\ew] a_p(\o)\\
= \Xi_d\ew(a_d(\o)-1)+\sum_{p\in D,\ p\neq d} \Xi_p\ew a_p(\o)  .
 \eaaa

 By (\ref{arho}), it follows that
\baaa\int_{-\pi}^\pi |W_d(\o)|^2 \WW(\o,\b_d,q,c)^2d\o\le r.
\label{W1}\eaaa
 By (\ref{arho1}), it follows that
\baa \int_{-\pi}^\pi |W_d\ew|^2 \WW(\o,\b,q,c)^2d\o = \int_{-\pi}^\pi |\Xi_d\ew|^2 d\o\le r.
\label{1}\eaa
 Hence $\yy_d\in \X_{\b}(q,c,2r) $ for all $d$.

In addition, it was shown above that $\w x_d(k)=\w x_0(k)$, for $k\le 0$ and
 $d>0$, and  $\w x_d(k)=\w x_0(k)$, for $k\ge 0$ and
 $d<0$.

 Let $\w x$
 be defined by (\ref{xPm}) with  $\yy_d$ defined above.
 By the definitions, it follows that   $\w x\in\PP_{m,\oo\b}(\qq,c,r)$.

Further, assume that $q>1$ is fixed and $c\to 0+$. Clearly, \baaa
 &&\|a_p\|_{L_2(-\pi,\pi)}\to 0,\quad
 \|A\|_{L_2(-\pi,\pi)}\to 1\quad \hbox{as}\quad c\to 0+ .
\label{aA}\eaaa
 It follows that  \baaa
 &&\|\w \YY_0\ew-\YY_0\ew\|_{L_2(-\pi,\pi)}\to 0,\\
 &&\|\w \YY_d\ew-\YY_d\ew\|_{L_2(-\pi,\pi)}\to 0\quad  \hbox{as}\quad c\to 0+ .
\label{aAA}\eaaa
It follows that \baa
 \|\w\yy_d-\yy_d\|_{\ell_2}\to 0\quad \hbox{as}\quad c\to 0.
\label{xdd1}\eaa
By the definitions, it follows that
\baa
\w\yy_d\left(\frac{l+d}{m}\right)-\yy_d\left(\frac{l+d}{m}\right)=\w x(l)-x(l)\label{xy}\eaa
for $l\in\ZZ$ such that $(l+d)/m\in\ZZ$. More precisely, (\ref{xy}) holds if  $d=0$, and it also holds if either $d\in \ZZ^+_1$ and $l\in\ZZ_0^+$,  or if $d\in \ZZ^0_{-1}$ and  $l\in\ZZ_0^-$.
For $k=(l+d)/m$ and $l=km-d$,  equation (\ref{xy}) can be rewritten as
\baaa
\w x(mk-d)-x(mk-d)=\w\yy_d(k)-\yy_d(k).
\eaaa
Finally, we obtain that  (\ref{xdd}) follows from (\ref{xdd1}) and (\ref{xy}). This completes the proof of Theorem \ref{ThDense}. $\Box$

\par
{\em Proof of Theorem \ref{ThRR}} follows from Lemma \ref{ThR} applied to the prediction of the corresponding subsequences. $\Box$

\noxxx{

{\em Proof of Theorem \ref{ThCTD}}.  The previous proof shows that $\ww x\in\ell_2$.
For $\ww X=\Z \ww x$,  we have that
\baaa
\ww x(k)=\frac{1}{2\pi}\int_{-\pi}^{\pi}
\ww X\left(e^{i\nu}\right) e^{i\nu k}d\nu\brea =\frac{\tau}{2\pi}\int_{-\pi/\tau}^{\pi/\tau}\ww X\left(e^{i\tau\o}\right) e^{i\o\tau k}d\o,
\eaaa
with the change of variables $\nu=\tau \o$. Let us define function $\ww F: i\R\to \C$ as  $\ww F(i\o)\defi\tau \ww X\left(e^{i \tau\o  }\right)$ for $\o\in\R$.
Then
\baaa
\ww x(k)=\frac{1}{2\pi}\int_{-\pi/\tau}^{\pi/\tau}
\ww F\left(i\o\right) e^{i\o\tau k}d\o.
\eaaa
 Since $\ww X\left(e^{i\cdot}\right)\in L_2(-\pi,\pi)$, this implies   that $\ww F\left(i\cdot\right)\in L_2(i\R)$.
The  sequence $\{\ww x(k)\}_{k\in \ZZ}$ represents the sequence of Fourier coefficients of $\ww F$ and defines  $\ww F$ uniquely. By Theorem \ref{ThDeg}, this sequence is uniquely defined
 by the sequence $\{\ww x(mk)\}_{k\in \ZZ}$.  Let $\ww f\defi \F \ww F$. Clearly,  $\ww f\in \BLOO$ and it is uniquely defined
 by the sequence $\{\ww x(mk)\}_{k\in \ZZ}$.
This completes the proof of Theorem \ref{ThCTD}.
$\Box$

{\em Proof of Corollary \ref{corr1}}.    We use notations from the proofs above.
The existences of the required $\{\w x _d\}_{-m+1}^{m-1}$, $\ww x$, and $\ww f$,  follows from Theorem \ref{ThDeg}-\ref{ThDense}.
We have for $F=\F  f$ that
\baaa
&&\|\ww F(i\cdot)-F(i\cdot)\|_{L_2(\R)}= \sqrt{\tau}\|\ww x-x\|_{\ell_2} \breakk \le
\sqrt{\tau}\sum_{d=-m+1}^{m-1}\|\w x_d-x_d\|_{\ell_2}\le  \sqrt{\tau}(2m-1) \e.
\eaaa
Since $F(i\o)=0$ and $\ww F(i\o)=0$ if $|\o|>\pi/\tau$,
it follows that \baaa
\|\ww F(i\cdot)-F(i\cdot)\|_{L_1(\R)}\le \sqrt{2\O}\|\ww F(i\cdot)-F(i\cdot)\|_{L_2(\R)}.
\eaaa
Combining these estimates, we obtain  that
 \baaa
 \|\ww f-f\|_{L_p(\R)}\le  2\pi\max(1,\sqrt{2\O})\|\ww F(i\cdot)-F(i\cdot)\|_{L_2(\R)}\brea \le
2\pi\max(1,\sqrt{2\O})\sqrt{\tau}(2m-1) \e .
 \eaaa
This completes the proof of  Corollary \ref{corr1}. $\Box$
\par
{\em Proof of Corollary \ref{ThRRC}.} It suffices to observe that $\ww x(k)=\ww f(t_k)$
satisfy the assumptions of  Corollary \ref{ThRR}. $\Box$

\par
 {\em Proof of Corollary \ref{corr2}.}
 We have that
 \baaa
&&\sup_{t\in\R}|f(t)-f_{\O}(t)|\le \frac{1}{2\pi}\int_{-\infty}^\infty |F(i\o)- F_{\O}(i\o)|d\o\breakk=\frac{1}{2\pi}\int_{\R\setminus [-\O,\O]} |F(i\o)|d\o
 \to 0\quad\\ &&\hbox{as}\quad \O\to +\infty,\ m\to+\infty.\eaaa
 Then statement (i) follows. The function $f_{\O}$ has the same properties as the function $f$ in Corollary \ref{corr1}. This completes  the proof of Corollary \ref{corr2}.  $\Box$
\section{Discussion and future developments} The paper shows that recoverability of sequences from their decimated subsequences
is associated with certain spectrum degeneracy of a new kind, and that a
sequences of a general kind can be approximated by sequences featuring this degeneracy.
   This is applied to sampling of continuous time band-limited functions.
The paper suggests a uniqueness result for  sampling  below the Nyquist rate of a class
of continuous time functions that is everywhere dense in the class of all band-limited functions.
(Theorems \ref{ThCTD} and Corollary \ref{corr1}), and establishes some  robustness of recovery
(Corollary \ref{ThRRC}). This means that the restriction on the sampling rate defined by the  Nyquist rate the Sampling Theorem is bypassed, in a certain sense. This was only the very first step in attacking  the problem; the  numerical implementation to data compression and recovery is quite challenging and there are many open questions. In particular, it involves the solution of ill-posed inverse problems. In addition, there are other possible extensions of this work that we will leave for the future research.
\begin{enumerate}
\xxxonly{\item
A theoretical problem arises: {\em How to detect a trace $x_0|_{k\le 0}$ that can be extended
into a process featuring degeneracy  described in Definition \ref{defDeg} for $d>0$?}
This is actually a non-trivial question even for $m=1$; see discussion Definition 1 in \cite{DX16b}
and discussion in \cite{Df,DX16b,D17b}.}
\item To cover applications related to  image processing, the approach of this paper has to be extended  on  2D sequences and functions $f:\R^2\to\R$.
This seems to be a quite difficult task; for instance, our approach is based on the predicting algorithm \cite{D12a} for 1D sequences, and it is unclear if it
can be extended on processes defined on multidimensional lattices.  Possibly, the setting from \citet{PM} can be used.
\item The result of this paper allows many acceptable modifications such as the following.
 \subitem -- The choice of mappings $\M_d$ allows many modifications;  for example,
 $x_d=\M_d x$ are defined such that  there exists $\t\ge m$ such that
$x_d(k)=x(k-d)$ for $k>\t$, without any restrictions on $\{x_d(k)\}_{k=1,..,\t}$.
\subitem --    Conditions of Theorem \ref{ThOdd} can be relaxed: instead of the condition that the spectrum  vanishes  on open sets $J_{m\nu_d}(\d)$, we can require that the  spectrum  vanishes only at  $R_{m\nu_d}$; however, the rate of vanishing at these points has to be sufficient similarly to \citet{D12a}. This is because
we use here the predicting algorithm based on the algorithm from  \citet{D12a} that does not  require that the spectrum of an underlying process is vanishing on an open subset of $\T$.
\subitem --  The sets $R_n$ of roots of order $n$ from $(-1)$  could be replaced by sets of roots from any point at $\TT$, with some  minor modifications.
\subitem --  The choice (\ref{wH}) of predictors presented in the proofs above is not unique. For example,
we could use  a predicting algorithm from \citet{D12b}  instead of
the algorithm based on \citet{D12a} used above.
Some examples of numerical experiments for the predicting algorithm based on the transfer function (\ref{wH}) \xxxonly{assumed by Step (R1)}  can be found in  \citet{D16} (for the case where $m=n=1$, in the notations of the present papers).
 \end{enumerate}

}
\xxxonly{}
\end{document}